\pdfoutput=1
\documentclass[10pt,journal]{IEEEtran}

\usepackage[T1]{fontenc}
\usepackage[utf8]{inputenc}
\usepackage{cite}
\usepackage{amsmath,amssymb,amsfonts,bm}
\usepackage{mathtools}
\usepackage{graphicx}
\usepackage{textcomp}
\usepackage{algorithm}
\usepackage{algorithmic}
\usepackage{tikz}
\usepackage[hidelinks]{hyperref}
\usepackage{orcidlink}

\hypersetup{
  pdftitle={Stability Transfer for Autonomous Systems via Higher-Order Taylor-Truncated Lyapunov Certificates},
  pdfauthor={Jose D. Hoyos, Zehui Lu, and Shaoshuai Mou},
  pdfsubject={Author's manuscript; IEEE Control Systems Letters, 2026},
  pdfkeywords={Lyapunov methods, Stability of nonlinear systems, Computational methods, Optimization, LMIs}
}

\DeclarePairedDelimiter{\norm}{\lVert}{\rVert}

\newtheorem{theorem}{Theorem}
\newtheorem{lemma}{Lemma}

\newtheorem{defn}{Definition}
\newtheorem{exmp}{Example}
\newtheorem{rem}{Remark}
\newtheorem{prop}{Proposition}

\newenvironment{proof}[1][Proof]
  {\begin{IEEEproof}[#1]}{\end{IEEEproof}}

\begin{document}

\def\BibTeX{{\rm B\kern-.05em{\sc i\kern-.025em b}\kern-.08em
    T\kern-.1667em\lower.7ex\hbox{E}\kern-.125emX}}

\title{Stability Transfer for Autonomous Systems via Higher-Order Taylor-Truncated Lyapunov Certificates}

\author{
Jose D. Hoyos\orcidlink{0000-0001-8768-115X}, \textit{Graduate Student Member, IEEE,}
Zehui Lu\orcidlink{0000-0002-7312-7075},
and Shaoshuai Mou\orcidlink{0000-0002-3698-4238}
\thanks{This is the author's manuscript of J.~D.~Hoyos,
Z.~Lu, and S.~Mou, ``Stability Transfer for Autonomous Systems via
Higher-Order Taylor-Truncated Lyapunov Certificates,''
\textit{IEEE Control Systems Letters}, vol.~10, pp.~1135--1140, 2026.
The published version is available at
\url{https://doi.org/10.1109/LCSYS.2026.3705005}.
\copyright~2026 IEEE. Personal use of this material is permitted.
Permission from IEEE must be obtained for all other uses, in any current
or future media, including reprinting/republishing this material for
advertising or promotional purposes, creating new collective works,
for resale or redistribution to servers or lists, or reuse of any
copyrighted component of this work in other works.}
\thanks{Jose D. Hoyos and Shaoshuai Mou are with the School of Aeronautics and Astronautics, Purdue University, West Lafayette, IN 47907, USA; \texttt{jhoyos@purdue.edu}, \texttt{mous@purdue.edu}}
\thanks{Zehui Lu is an independent researcher; \texttt{zehuilu789@gmail.com}}
}

\maketitle
\thispagestyle{plain}

\begin{abstract}
We develop analytical sufficient conditions for certifying the local asymptotic stability of nonlinear autonomous systems based on Lyapunov functions derived from their finite-order Taylor polynomial approximations. This approach translates the stability analysis problem from the original nonlinear dynamics to a more computationally tractable polynomial domain. The derived conditions allow for the explicit calculation of a guaranteed inner approximation of the region of attraction ($\mathrm{ROA}$). We show how sum-of-squares (SOS) programming and grid search can be used to systematically search for polynomial Lyapunov functions that yield a large certified threshold radius and extract a certified invariant set inner approximation of the $\mathrm{ROA}$ of the nonpolynomial original system. The framework is then applied to a control tuning problem: finding reduced-effort Linear Quadratic Regulator (LQR) controllers that ensure that a given initial state of interest belongs to the certified $\mathrm{ROA}$ of the nonpolynomial closed loop system. The effectiveness of the overall methodology is demonstrated using a pendulum with nonlinear damping and a nonlinear control system.
\end{abstract}


\section{Introduction}
\label{sec:introduction}

\IEEEPARstart{V}{erifying} the stability of nonlinear autonomous systems is a central problem in control \cite{khalil2002nonlinear,meiss2007differential}. The Lyapunov direct method provides a general framework, but constructing suitable Lyapunov functions is often non-systematic \cite{khalil2002nonlinear,meiss2007differential}. For polynomial vector fields, sum-of-squares (SOS) programming enables automated searches for polynomial Lyapunov functions by casting polynomial nonnegativity as semidefinite programs \cite{parrilo2000structured}, with applications to optimal control \cite{ichihara2009optimal}, stability verification \cite{TOPCU20082669}, and region-of-attraction ($\mathrm{ROA}$) estimation \cite{8959232}. A common strategy is to rely on local approximations to recover a tractable analysis framework.

A classical approximation uses the Jacobian linearization, which can certify local stability on an unspecified sufficiently small neighborhood \cite{khalil2002nonlinear,meiss2007differential}. In this paper, we instead leverage higher order Taylor approximations to obtain a polynomial analysis framework while rigorously relating stability of the Taylor polynomial to stability of the original nonpolynomial dynamics on an explicit region. Our main result provides sufficient conditions under which Lyapunov decay certified for the Taylor polynomial transfers to the true system by bounding the Taylor remainder and requiring the polynomial decay to dominate this bound. As a consequence, the proposed approach certifies local asymptotic stability and yields a computable inner approximation of the $\mathrm{ROA}$.

To make the framework constructive, we develop a SOS computational pipeline to search for polynomial functions satisfying the proposed conditions, coupled with a grid search. We illustrate the approach in a controller tuning task by selecting minimum-effort LQR gains over a grid of candidates that certify that a prescribed initial condition lies within the $\mathrm{ROA}$ of the nonpolynomial closed loop system.

Prior work has shown that nonpolynomial dynamics can be studied through polynomial surrogates. For example, \cite{mastellone2004nonlinear} used polynomial approximations with user-specified error bounds on prescribed regions, together with perturbation arguments and S-procedure-type reformulations based on auxiliary variables and inequality constraints. Later, \cite{chesi2009estimating} proposed an LMI-based domain-of-attraction analysis for a structured class of systems $\dot{\bm x}=\bm f(\bm x)+\sum_i \bm g_i(\bm x)\xi_i(x_{a_i})$, in which each nonpolynomial term depends on a single state coordinate. Their method combines truncated Taylor expansions with a polytopic description of the remainder and verifies the resulting polynomial inequalities at the vertices of the remainder polytope. By contrast, this paper addresses general smooth autonomous systems through a Taylor truncation of the full vector field and a multivariate remainder bound. The main result provides an explicit computable threshold radius guaranteeing that Lyapunov decrease for the truncated model transfers to the true system, yielding a certified positively invariant connected inner approximation of the region of attraction. The SOS-based procedure is used both for analysis and for controller 
tuning, searching over LQR weights and associated Lyapunov certificates 
to verify that a prescribed initial condition lies within the 
$\mathrm{ROA}$ of the nonpolynomial closed-loop system. An additional practical advantage is the direct computational setup: one constructs the Taylor polynomial, bounds the local Taylor remainder, and searches for a polynomial Lyapunov certificate via SOS programming, avoiding the involved auxiliary-variable formulations and nested-loop setups often needed in nonpolynomial analysis.

\noindent\textbf{\textit{Notations.}} Bold symbols denote vectors (e.g., $\bm{x}, \bm{y} \in \mathbb{R}^n$), and bold functions denote vector-valued maps (e.g., $\bm{f} \colon \mathbb{R}^n \mapsto \mathbb{R}^s$, $\bm{y}\! =\! \bm{f}(\bm{x})$). The Euclidean norm of $\bm{x} \!\in \!\mathbb{R}^n$ is $\norm{\bm{x}}_2$. For a radius $r\!>\!0$, the open Euclidean ball centered at $\bm a \in \mathbb R^n$ is
$B_r(\bm a)\!=\!\{\bm x\!\in\!\mathbb R^n\!:\!\|\bm x-\bm a\|_2\!<\!r\}$ and the closed ball is denoted as $\overline B_r(\bm a)$, where the overline means the closure of the set. For a point $\bm x\in\mathbb R^n$ and a set $S\subset\mathbb R^n$, the distance from $\bm x$ to $S$ is $\mathrm{dist}(\bm x,S)=\inf_{\bm y\in S}\|\bm x-\bm y\|_2.$ For a symmetric matrix $Q\!\in\!\mathbb R^{n\times n}$, we write $Q\!\succ\!0$ to denote that $Q$ is positive definite. For a $\mathcal{C}^k$ map $\bm f=[f_1,\ldots,f_n]^\top:\mathbb{R}^n\mapsto\mathbb{R}^n$, the $k$-linear operator $\nabla^k\bm f(\bm a):(\mathbb{R}^n)^k \mapsto \mathbb{R}^n$ denotes the $k$-th partial-derivative tensor with entries $\big(\nabla^k\bm f(\bm a)\big)_{\ell,i_1,\ldots,i_k}=\frac{\partial^k f_\ell}{\partial x_{i_1}\cdots\partial x_{i_k}}(\bm a)$, symmetric in $(i_1,\ldots,i_k)$. For $\bm v_1,\ldots,\bm v_k\in\mathbb R^n$,
$\nabla^k\bm f(\bm a)(\bm v_1,\ldots,\bm v_k)$ denotes the corresponding multilinear evaluation. In particular, we use the shorthand $\nabla^{k}\bm f(\bm a)(\bm x)^{(k)}=\nabla^{k}\bm f(\bm a)(\bm x,\ldots,\bm x)\in\mathbb R^n$. When $\bm f$ is $\mathcal{C}^1$, we write $J\bm f(\bm a)=\nabla \bm f(\bm a)$ for its Jacobian. For a scalar $V:\mathbb{R}^n\mapsto\mathbb{R}$, $\nabla V$ is its gradient and the Lie derivative along $\bm f$ is $\mathcal{L}_f V(\bm x)=\nabla V(\bm x)^\top \bm f(\bm x)$. The induced operator norm is $\|\nabla^{k}\bm f(\bm a)\|_{\mathrm{op}}=\sup_{\|\bm x_i\|_2=1}\,\big\|\nabla^{k}\bm f(\bm a)(\bm x_1,\ldots,\bm x_k)\big\|_2.$

\section{Problem Formulation}
\label{sec:ProblemFormulation}

This section introduces the problem setting and preliminaries. We recall standard notions of asymptotic stability and Lyapunov functions for autonomous systems, and then formalize the objective: certify stability of a nonlinear autonomous system via analysis of a finite order polynomial Taylor approximation with explicit estimate of the $\mathrm{ROA}$.

\subsection{Preliminaries}
\label{sec:Preliminaries}

We study autonomous systems $\dot{\bm x}=\bm f(\bm x)$ with time $t\in\mathbb{R}_{\ge 0}$ and the state vector $\bm x(t)\in\mathcal X\subset\mathbb{R}^n$, where $\mathcal X$ is open and $\bm f:\mathcal X\mapsto\mathbb{R}^n$ is locally Lipschitz, ensuring local existence and uniqueness for any $\bm x(0)\in\mathcal X$ \cite{khalil2002nonlinear,meiss2007differential}.

\begin{defn}[Local and Asymptotic Stability \cite{khalil2002nonlinear}]\label{def:Stability_Definitions}
A point $\bm{a} \in \mathcal{X}$ s.t. $\bm{f}(\bm{a}) = \bm{0}$ of the system $\dot{\bm{x}} = \bm{f}(\bm{x})$, is called:
\begin{enumerate}
    \item Locally Stable: if for any $\varepsilon > 0$, there exists a $\delta > 0$ such that if $\norm{\bm{x}(0) - \bm{a}}_2 < \delta$, then $\norm{\bm{x}(t) - \bm{a}}_2 < \varepsilon$ for all $t \ge 0$.
    
    \item Locally Asymptotically Stable: if it is locally stable and, in addition, it is locally attractive, i.e., there exists a $\delta_a > 0$ such that if $\norm{\bm{x}(0) - \bm{a}}_2 < \delta_a$, then $\lim_{t \to \infty} \bm{x}(t) = \bm{a}$.
\end{enumerate}
The set of initial states $\bm{x}(0)$ that converge to $\bm{a}$ as $t \to \infty$ is the region of attraction $\mathrm{ROA}(\bm a)$. Any set $S \subset$ $\mathrm{ROA}(\bm a)$ such that $\bm a \in S$ is called an inner approximation of $\mathrm{ROA}(\bm a)$. 
\end{defn}

\begin{defn}[Lyapunov Function \cite{khalil2002nonlinear}]\label{def:local_LF}
A $\mathcal{C}^1$ function $V: \Omega \mapsto \mathbb{R}$, defined on an open set $\Omega \subset \mathcal{X}$ containing the equilibrium point $\bm{a}$, is a local Lyapunov function if:
\begin{enumerate}
    \item Positive Definiteness: $V(\bm{a}) \!=\! 0$, $V(\bm{x})\! > \!0,\forall \bm{x}\! \in\! \Omega \!\setminus \!\{\bm{a}\}$.
    \item Lie-derivative Negative Semidefiniteness: $\mathcal{L}_f V(\bm{x}) \!=\! \nabla V(\bm{x})^{\!\top}\! \bm{f}(\bm{x})$, is negative semidefinite, $\mathcal{L}_f V(\bm{x})\! \le\! 0,\forall\bm{x}\!\in \!\Omega$.
\end{enumerate}
If the Lie-derivative is negative definite (i.e., $\mathcal{L}_f V(\bm{x}) < 0$ for all $\bm{x} \in \Omega \setminus \{\bm{a}\}$), $V$ is a strict local Lyapunov function.
\end{defn}

\begin{lemma}[Lyapunov Theorem for Local Stability \cite{khalil2002nonlinear}]\label{thm:lf_implies_stability}
If a Lyapunov function $V: \Omega \mapsto \mathbb{R}$ exists, then the equilibrium point $\bm{a}$ of $\dot{\bm{x}} = \bm{f}(\bm{x})$ is locally stable. If $V(\bm x)$ is a strict Lyapunov function, then $\bm{a}$ is locally asymptotically stable.
\end{lemma}

\begin{defn}[Taylor Polynomial of $p$th-order]\label{def:Taylor_polynomial}
Let $\mathcal X\subset\mathbb R^{n}$ be an open set, and let the vector field $\bm f:\mathcal X\mapsto\mathbb R^{n}$ be of class $\mathcal C^{p+1}$ for some $p\ge0$. For a fixed point $\bm a\!\in\!\mathcal X$ and a displacement $\bm h=\bm x-\bm a \in\mathbb R^{n}$ such that the line segment $\{\bm a+\theta\bm h:0\le \theta\le1\}$ is contained in $\mathcal X$ (e.g., if $\mathcal X$ is convex, $\bm a \!\in\! \mathcal{X}$ and $\bm a +\bm h \!\in\! \mathcal X$ suffice), the $p$th-order Taylor polynomial of $\bm f$ centered at $\bm a$ is defined as
\begin{equation}\label{eq:taylor_poly}
    \bm f_p(\bm a+\bm h)=\sum_{k=0}^{p}\frac1{k!}\,
    \nabla^{\,k}\bm f(\bm a)(\bm h)^{(k)}.
\end{equation}
\end{defn}

\subsection{Problem of Interest}
\label{sec:ProblemofInterest}

Given $\dot{\bm x}=\bm f(\bm x)$ with equilibrium point $\bm x =\bm a$, assume that the
$p$th-order Taylor model $\dot{\bm x}=\bm f_p(\bm x)$ is locally
asymptotically stable at $\bm x =\bm a$ (Definition~\ref{def:Stability_Definitions})
and admits a polynomial Lyapunov function $V$ on $\Omega$
(Definition~\ref{def:local_LF}). We seek sufficient conditions under which
$\bm x =\bm a$ is also locally asymptotically stable for the original system
$\dot{\bm x}=\bm f(\bm x)$, and obtain an explicit inner approximation of
$\mathrm{ROA}(\bm a)$. We also develop SOS-based methods to search for such certificates and enlarge the certified radius of the resulting inner approximation of the $\mathrm{ROA}(\bm a)$.

\section{Main Result}
\label{sec:MainResult}

This section develops sufficient conditions that transfer asymptotic stability from the $p$th-order Taylor truncation $\bm f_p$ to the original dynamics $\bm f$. The key idea is to bound the Taylor remainder $\bm\rho_{p+1}(\bm x)=\bm f(\bm x)-\bm f_p(\bm x)$ and ensure that the Lyapunov decrease along the truncated dynamics is large enough to dominate the effect of this remainder on an explicit region. We begin with a
bound on this remainder.

\begin{lemma}[Bound for Vector Field Taylor Remainder]\label{thm:lagrange}
Let $\mathcal X \subset \mathbb{R}^n$ be open, $\bm{f} \colon \mathcal X \mapsto \mathbb{R}^n$ be $\mathcal{C}^{p+1}$ ($p \geq 0$), and assume $\exists m > 0$ such that
$\norm{\nabla^{p+1}\bm{f}(\bm{x})}_{\mathrm{op}} \leq m$ for all $\bm{x} \!\in\! \mathcal X$. Fix $\bm{a} \!\in\! \mathcal X$ and $\bm{h} \!\in\! \mathbb{R}^n$ with $\{\bm{a} + \theta\bm{h} \colon\! 0 \leq \theta\leq 1\} \subset \mathcal X$ (if $\mathcal X$ is convex, $\bm{a} \!\in\! \mathcal X$ and $\bm{a} + \bm{h} \!\in\! \mathcal X$ suffice). Define the remainder
\[
\bm{\rho}_{p+1}(\bm{h}) = \bm{f}(\bm{a} + \bm{h}) - \sum_{k=0}^p \frac{1}{k!}\nabla^{k}\bm{f}(\bm{a})(\bm{h})^{(k)},\quad \text{then}
\]
\[
\norm{\bm{\rho}_{p+1}(\bm{h})}_2 \leq m \frac{\norm{\bm{h}}_2^{p+1}}{(p+1)!}.
\]
\end{lemma}

\begin{proof}
See Theorem 5.6.2 of \cite{cartan1971differential} and p.~492 of \cite{lang2005undergraduate}.
\end{proof}

\begin{rem}[Local Taylor Remainder Bound]\label{rem:compact_bound}
The global bound $m$ in Lemma~\ref{thm:lagrange} may fail on unbounded domains. For local analysis around a fixed point $\bm a\in\mathcal X$, since $\mathcal X$ is open, choose $r>0$ such that $\overline B_r(\bm a)\subset\mathcal X$, and set
$\Omega=B_r(\bm a)$. Then $\overline{\Omega}$ is compact by the Heine--Borel theorem \cite[p.~40]{Rudin1976}. Since
$\bm f\in\mathcal C^{p+1}(\mathcal X;\mathbb R^n)$, the map $\bm x\mapsto \|\nabla^{p+1}\bm f(\bm x)\|_{\mathrm{op}}$ is continuous on
$\overline{\Omega}$ and attains a finite maximum $m_{\overline{\Omega}}:=\max_{\bm x\in\overline{\Omega}}
\|\nabla^{p+1}\bm f(\bm x)\|_{\mathrm{op}}<\infty$. Hence, for every $\bm h$ such that the segment
$\{\bm a+\theta\bm h:0\leq\theta\leq1\}$ is contained in $\overline{\Omega}$, the Taylor remainder satisfies
$\|\bm\rho_{p+1}(\bm h)\|_2\leq m_{\overline{\Omega}}\|\bm h\|_2^{p+1}/(p+1)!$.
\end{rem}

We now present sufficient conditions ensuring that the Lyapunov decrease certified along the Taylor-truncated dynamics,
$\nabla V(\bm x)^\top \bm f_p(\bm x)<0$, is preserved for the original system $\dot{\bm x}=\bm f(\bm x)$ on an explicit ball around $\bm a\in\mathcal X$, and that Lyapunov sublevel sets in that ball lie in $\mathrm{ROA}(\bm a)$.

\begin{theorem}[Sufficient Conditions for Stability Transfer]\label{thm:lie-decay-region}
Let $\mathcal X\subset\mathbb R^{n}$ be nonempty and open, and let
$\bm f\in\mathcal C^{p+1}(\mathcal X;\mathbb R^{n})$ for some $p\ge0$.
Fix $\bm a\in\mathcal X$ such that $\bm f(\bm a)=\bm0$, and write the
$p$th-order Taylor expansion of $\bm f$ at $\bm a$ as
$\bm f(\bm x)=\bm f_p(\bm x)+\bm\rho_{p+1}(\bm x)$, where $\bm f_p$ denotes
the $p$th-order Taylor polynomial of $\bm f$ at $\bm a$
(Definition~\ref{def:Taylor_polynomial}, equation~\eqref{eq:taylor_poly}),
and $\bm\rho_{p+1}$ is the remainder (Lemma~\ref{thm:lagrange}). Suppose there exists a bounded open set $\Omega\subset\mathcal X$ with $\bm a\in\Omega$ and $\overline{\Omega}\subset\mathcal X$, and a constant $m>0$ such that $\|\nabla^{p+1}\bm f(\bm x)\|_{\mathrm{op}}\le m$ $\forall\bm x\in\overline{\Omega}$. Define $k_p=m/(p+1)!$. Suppose there exists a
polynomial $V:\Omega\mapsto\mathbb R$, with $V(\bm a)=0$ and
$V(\bm x)>0$ $\forall\bm x\in\Omega\setminus\{\bm a\}$, such that:
\begin{enumerate}\renewcommand{\labelenumi}{(\roman{enumi})}
    \item\label{b1} $\nabla V(\bm x)^{\!\top}\bm f_{p}(\bm x)\le0$ for all $\bm x\in\Omega$;

    \item\label{b2} $\nabla V(\bm x)\neq0$ for all $\bm x\in\Omega\setminus\{\bm a\}$;

    \item\label{dec-2} There exist $\mu>0$, $\delta>0$, and $0\le c\le p+1$ such that, for all
    $\bm x\in(B_\delta(\bm a)\cap\Omega)\setminus\{\bm a\}$,
    \begin{equation}\label{eqdec-2}
    -\nabla V(\bm x)^{\!\top}\bm f_{p}(\bm x)
    \ge
    \mu\,\|\nabla V(\bm x)\|_2\,\|\bm x-\bm a\|_2^c,
    \end{equation}
    and, if $c=p+1$, assume additionally that $\mu>k_p$.
\end{enumerate}

Define $\bar r=\min\{(\mu/k_p)^{1/(p+1-c)},\delta,\mathrm{dist}(\bm a,\partial\Omega)\}$ if $c<p+1$, and
$\bar r=\min\{\delta,\mathrm{dist}(\bm a,\partial\Omega)\}$ if $c=p+1$.
Then, for every $r\in(0,\bar r)$, $V$ is a strict local Lyapunov function for
the original system $\dot{\bm x}=\bm f(\bm x)$ on $B_r(\bm a)$. In particular,
$\bm a$ is a locally asymptotically stable equilibrium of
$\dot{\bm x}=\bm f(\bm x)$. Moreover, define
$\gamma_r^\star=\min_{\|\bm x-\bm a\|_2=r}V(\bm x)$,
$S_\gamma=\{\bm x\in\Omega:V(\bm x)\le\gamma\}$, and let
$\mathcal L_\gamma$ be the connected component of $S_\gamma$ containing
$\bm a$. Then $\gamma_r^\star>0$, and every
$\gamma\in(0,\gamma_r^\star)$ satisfies
$\mathcal L_\gamma\subset B_r(\bm a)$, $\mathcal L_\gamma$ is positively
invariant, and $\mathcal L_\gamma\subset\mathrm{ROA}(\bm a)$.
\end{theorem}

\begin{proof}
Fix any $r\in(0,\bar r)$. Regardless of whether $c<p+1$ or $c=p+1$,
$\bar r$ is defined so that $\bar r\le\delta$ and
$\bar r\le\mathrm{dist}(\bm a,\partial\Omega)$. Hence
$B_r(\bm a)\subset B_\delta(\bm a)\cap\Omega$. Moreover, for every
$\bm x\in B_r(\bm a)$, the segment
$\{\bm a+\theta(\bm x-\bm a):0\le\theta\le1\}$ lies in
$B_r(\bm a)$, and hence in $\Omega\subset\mathcal X$. Therefore, by
Lemma~\ref{thm:lagrange} with the assumed bound $m$ on $\overline{\Omega}$,
$\|\bm\rho_{p+1}(\bm x)\|_2\le
k_p\|\bm x-\bm a\|_2^{p+1}$ $\forall\bm x\in B_r(\bm a)$.

\textit{Case 1:} \(c<p+1\). Let
$r_1=(\mu/k_p)^{1/(p+1-c)}$. By the definition of $\bar r$, we have
$r<r_1$. For every \(\bm x\in B_r(\bm a)\setminus\{\bm a\}\), we have
$\|\bm x-\bm a\|_2<r<r_1$. This implies
$\|\bm x-\bm a\|_2^{p+1-c}<r_1^{p+1-c}=\mu/k_p$, and thus
$k_p\|\bm x-\bm a\|_2^{p+1-c}<\mu$. Since
$\|\bm x-\bm a\|_2^c>0$ for $\bm x\neq\bm a$, we obtain
\begin{equation}\label{eq:ineq}
k_p\|\bm x-\bm a\|_2^{p+1}<\mu\|\bm x-\bm a\|_2^c.
\end{equation}
By the Cauchy--Schwarz inequality, the local remainder estimate above, and
assumption~\ref{dec-2},
\begin{align*}
&\nabla V(\bm x)^{\!\top} \bm f(\bm x)
  \le \nabla V(\bm x)^{\!\top} \bm f_{p}(\bm x)
      + \big| \nabla V(\bm x)^{\!\top} \bm{\rho}_{p+1}(\bm x)\big| \\
  &\le \nabla V(\bm x)^{\!\top} \bm f_{p}(\bm x)
      + \|\nabla V(\bm x)\|_2  \|\bm{\rho}_{p+1}(\bm x)\|_2  \\
  &\le \nabla V(\bm x)^{\!\top} \bm f_{p}(\bm x)
      + \|\nabla V(\bm x)\|_2  k_{p} \|\bm x - \bm a\|_2^{p+1} \\
  &\le  -\mu\,\|\nabla V(\bm x)\|_2\,\|\bm x-\bm a\|^c_2
       + k_{p}\,\|\nabla V(\bm x)\|_2\,\|\bm x-\bm a\|^{p+1}_2  \\
  &=\|\nabla V(\bm x)\|_2 (-\mu\|\bm x-\bm a\|^c_2 +k_{p}\|\bm x-\bm a\|^{p+1}_2).
\end{align*}
Using \eqref{eq:ineq}, the term
$-\mu\|\bm x-\bm a\|^c_2+k_{p}\|\bm x-\bm a\|^{p+1}_2<0$.
By assumption~\ref{b2}, $\|\nabla V(\bm x)\|_2>0$ for
$\bm x\neq\bm a$. Therefore, $\nabla V(\bm x)^{\!\top}\bm f(\bm x)<0$ for all
$\bm x\in B_r(\bm a)\setminus\{\bm a\}$.

\textit{Case 2:} \(c=p+1\). For any
$\bm x\in B_r(\bm a)\setminus\{\bm a\}$, using the local remainder estimate
above and assumption~\ref{dec-2},
\begin{align*}
\mathcal{L}_f V(\bm x)
&\le
\|\nabla V(\bm x)\|_2
(-\mu\|\bm x-\bm a\|^{p+1}_2
+k_{p}\|\bm x-\bm a\|^{p+1}_2)\\
&=
\|\nabla V(\bm x)\|_2\|\bm x-\bm a\|^{p+1}_2(k_p-\mu).
\end{align*}
Since $\mu>k_p$, the term $(k_p-\mu)$ is strictly negative. By
assumption~\ref{b2} and $\bm x\neq\bm a$,
$\|\nabla V(\bm x)\|_2\|\bm x-\bm a\|_2^{p+1}>0$. Thus,
$\nabla V(\bm x)^{\!\top}\bm f(\bm x)<0$ on
$B_r(\bm a)\setminus\{\bm a\}$.

In both cases, and noting that $\mathcal L_fV(\bm a)=0$, we have constructed a
radius $r>0$ such that $V(\bm x)$ is positive definite on $B_r(\bm a)$ and
$\mathcal L_fV(\bm x)$ is negative definite on
$B_r(\bm a)\setminus\{\bm a\}$. This satisfies the conditions for $V$ to be a
strict local Lyapunov function (Definition~\ref{def:local_LF}). By
Lemma~\ref{thm:lf_implies_stability}, the existence of a strict local Lyapunov
function implies that the equilibrium $\bm a$ is locally asymptotically stable
for $\dot{\bm x}=\bm f(\bm x)$.

Now consider $\gamma_r^\star=\min_{\|\bm x-\bm a\|_2=r}V(\bm x)$. Since
$r<\mathrm{dist}(\bm a,\partial\Omega)$, the sphere
$\{\|\bm x-\bm a\|_2=r\}$ is compact and contained in $\Omega$. Since $V$ is
continuous and positive definite, $\gamma_r^\star>0$. Fix
$\gamma\in(0,\gamma_r^\star)$, and let
$S_\gamma=\{\bm x\in\Omega:V(\bm x)\le\gamma\}$ and
$\mathcal L_\gamma$ be the connected component of $S_\gamma$ containing
$\bm a$. We claim that $\mathcal L_\gamma\subset B_r(\bm a)$. If not, then
since $\mathcal L_\gamma$ is connected and contains $\bm a$, it must intersect
the sphere $\{\|\bm x-\bm a\|_2=r\}$. But on that sphere,
$V(\bm x)\ge\gamma_r^\star>\gamma$, contradicting
$\mathcal L_\gamma\subset S_\gamma$.

Next, $\mathcal L_\gamma$ is positively invariant. Let $\bm x(t)$ be a
trajectory with $\bm x(0)\in\mathcal L_\gamma$. While
$\bm x(t)\in B_r(\bm a)$, one has $\mathcal L_fV(\bm x(t))\le0$, hence
$V(\bm x(t))\le V(\bm x(0))\le\gamma$. Therefore the trajectory cannot cross
the sphere $\{\|\bm x-\bm a\|_2=r\}$, because there
$V\ge\gamma_r^\star>\gamma$. Thus $\bm x(t)\in B_r(\bm a)$ for all
$t\ge0$, and hence $\bm x(t)\in S_\gamma$ for all $t\ge0$. Since
$t\mapsto\bm x(t)$ is continuous, the image $\bm x([0,\infty))$ is a connected
subset of $S_\gamma$ containing $\bm x(0)\in\mathcal L_\gamma$. By definition
of $\mathcal L_\gamma$ as the connected component of $S_\gamma$ containing
$\bm a$, it follows that $\bm x([0,\infty))\subset\mathcal L_\gamma$.

Now $S_\gamma\cap\overline B_r(\bm a)$ is closed in the compact
$\overline B_r(\bm a)$, hence compact. Since $\mathcal L_\gamma$ is a
connected component of $S_\gamma\cap\overline B_r(\bm a)$, it is closed in
$S_\gamma\cap\overline B_r(\bm a)$ and then compact. Since
$\mathcal L_\gamma$ is compact and positively invariant, and
$\mathcal L_fV(\bm x)<0$ $\forall\bm x\in\mathcal L_\gamma\setminus\{\bm a\}$, the set
$\{\bm x\in\mathcal L_\gamma:\mathcal L_fV(\bm x)=0\}$ is $\{\bm a\}$. Hence,
by LaSalle's principle~\cite[Thm.~4.4]{khalil2002nonlinear}, every trajectory
starting in $\mathcal L_\gamma$ converges to $\bm a$. Therefore,
$\mathcal L_\gamma\subset\mathrm{ROA}(\bm a)$.
\end{proof}

\begin{prop}[Explicit Lower Bound for $\gamma_r^\star$]\label{prop:gamma-lower-bound}
Let $V$ satisfy the hypotheses of Theorem~\ref{thm:lie-decay-region}, and let
$\gamma_r^\star$ and $\mathcal L_\gamma$ be defined as in that theorem. Write
the translated polynomial $V(\bm a+\bm z)$ as the finite sum of its homogeneous
components,
$V(\bm a+\bm z)=\sum_{d=d_0}^{D}V_d(\bm z)$, where each $V_d$ is homogeneous of
degree $d$, and $d_0\ge2$ is the first index such that $V_{d_0}\neq0$. This
decomposition is available for any polynomial $V$ and does not require $V$
itself to be homogeneous. If, in addition, $V_{d_0}$ is positive definite,
then, with
$m_{d_0}=\min_{\|\bm z\|_2=1}V_{d_0}(\bm z)$ and
$b_d=\sum_{|\alpha|=d}|c_{d,\alpha}|$ when
$V_d(\bm z)=\sum_{|\alpha|=d}c_{d,\alpha}\bm z^\alpha$, one has
$\gamma_r^\star\ge\underline\gamma(r)=m_{d_0}r^{d_0}-\sum_{d=d_0+1}^{D}b_dr^d$.
Consequently, if $\underline\gamma(r)>0$, then every
$\gamma\in(0,\underline\gamma(r))$ satisfies
$\mathcal L_\gamma\subset\mathrm{ROA}(\bm a)$.
\end{prop}

\begin{proof}
Since $V_{d_0}$ is homogeneous and positive definite,
$V_{d_0}(\bm z)\ge m_{d_0}\|\bm z\|_2^{d_0}$. For each $d>d_0$, write
$V_d(\bm z)=\sum_{|\alpha|=d}c_{d,\alpha}\bm z^\alpha$ and
$b_d=\sum_{|\alpha|=d}|c_{d,\alpha}|$, so that
$|V_d(\bm z)|\le b_d\|\bm z\|_2^d$. Therefore,
$V(\bm a+\bm z)\ge m_{d_0}\|\bm z\|_2^{d_0}
-\sum_{d=d_0+1}^{D}b_d\|\bm z\|_2^d$. Evaluating on
$\|\bm z\|_2=r$ yields
\[
\gamma_r^\star
=
\min_{\|\bm z\|_2=r}V(\bm a+\bm z)
\ge
m_{d_0}r^{d_0}
-
\sum_{d=d_0+1}^{D}b_dr^d
=
\underline\gamma(r).
\]
Hence, if $\underline\gamma(r)>0$, then every
$\gamma\in(0,\underline\gamma(r))$ satisfies $\gamma<\gamma_r^\star$, and
therefore $\mathcal L_\gamma\subset\mathrm{ROA}(\bm a)$ by
Theorem~\ref{thm:lie-decay-region}.
\end{proof}

\begin{rem} Note that the radius $\bar r$ is not maximal. Also, if $V(\bm x)=(\bm x-\bm a)^\top P(\bm x-\bm a),$ with $P=P^\top\succ0$, then $d_0=2$, $m_2=\lambda_{\min}(P)$, and $b_d=0$ for all $d\ge3$. Therefore, the explicit lower bound becomes $\gamma_r^\star  =\underline \gamma (r)=\lambda_{\min}(P)\,r^2.$ Finally, note that the ball $B_{r}(\bm a)$ may or may not be inside $\mathrm{ROA}(\bm a)$, but $\mathcal L_\gamma\subset B_{r}(\bm a)$ and $\mathcal L_\gamma\subset \mathrm{ROA}(\bm a).$
\end{rem}

\begin{figure}
\centering
\resizebox{\columnwidth}{!}{%
\begin{tikzpicture}[scale=0.5,>=stealth]
\tikzset{
  every node/.style = {font=\small}
}

\draw[black, thick] (-4,-4) rectangle (4,4);
\node[above right] at (3,4) {$\mathbb{R}^n$};

\fill[orange!10] plot[smooth cycle] coordinates {
  (-3.5,1.5) (-2.5,3) (2,3.5) (3.5,1.5) (3,-1.5) (0.5,-3.5) (-2.5,-2.5)};
\draw[orange, thick, dashed] plot[smooth cycle] coordinates {
  (-3.5,1.5) (-2.5,3) (2,3.5) (3.5,1.5) (3,-1.5) (0.5,-3.5) (-2.5,-2.5)};
\node[orange] at (-3,3.2) {$\mathcal X$};

\fill[gray!30] plot[smooth cycle] coordinates {
  (-2.5,1) (-1.5,2) (1,2.5) (2.5,1) (2,-1) (0,-2.5) (-1.5,-1.5)};
\draw[black, thick, dashed] plot[smooth cycle] coordinates {
  (-2.5,1) (-1.5,2) (1,2.5) (2.5,1) (2,-1) (0,-2.5) (-1.5,-1.5)};
\node at (-1.5,2.5) {$\Omega$};

\coordinate (a) at (0,0);

\coordinate (boundary) at (-1.7,-1.);
\draw[->, thick] (a) -- (boundary) node[midway, above left] {$r_2$};
\node[below left] at (boundary) {$\partial\Omega$};

\draw[->, thick, red] (a) -- (45:3.5)
      node[midway, above right, xshift=15pt, yshift=7pt] {$\delta$};

\fill[green!25] plot[smooth cycle] coordinates {
  (-1.0,0.2) (-0.7,1.0) (0.2,1.15) (0.9,0.8) (1.0,-0.2) (0.45,-1.0) (-0.5,-0.95) (-1.0,-0.3)};
\draw[green!60!black, thick] plot[smooth cycle] coordinates {
  (-1.0,0.2) (-0.7,1.0) (0.2,1.15) (0.9,0.8) (1.0,-0.2) (0.45,-1.0) (-0.5,-0.95) (-1.0,-0.3)};
\node[green!50!black] at (0.45,0.6) {$\mathcal L_\gamma$};


\fill (a) circle (3pt) node[below] {$\bm a$};

\draw[blue, thick, dashed] (a) circle (2);
\node[blue] at (0,1.5) {$B_{r}(\bm a)$};

\node at (-1,4.5) {\textbf{Case: $\delta > r_2$ }};

\begin{scope}[xshift=9cm]
\draw[black, thick] (-4,-4) rectangle (4,4);
\node[above right] at (3,4) {$\mathbb{R}^n$};

\fill[orange!10] plot[smooth cycle] coordinates {
  (-3.5,1.5) (-2.5,3) (2,3.5) (3.5,1.5) (3,-1.5) (0.5,-3.5) (-2.5,-2.5)};
\draw[orange, thick, dashed] plot[smooth cycle] coordinates {
  (-3.5,1.5) (-2.5,3) (2,3.5) (3.5,1.5) (3,-1.5) (0.5,-3.5) (-2.5,-2.5)};
\node[orange] at (-3,3.2) {$\mathcal X$};

\fill[gray!30] plot[smooth cycle] coordinates {
  (-3.2,1.3) (-2.3,2.3) (1.3,2.8) (3.2,1.3) (2.7,-1.3) (0,-3.2) (-2.3,-1.8)};
\draw[black, thick, dashed] plot[smooth cycle] coordinates {
  (-3.2,1.3) (-2.3,2.3) (1.3,2.8) (3.2,1.3) (2.7,-1.3) (0,-3.2) (-2.3,-1.8)};
\node at (-2.2,1.3) {$\Omega$};

\coordinate (a2) at (0,0);
\coordinate (boundary2) at (-1.2*2.05,-0.7*2.05);
\draw[->, thick] (a2) -- (boundary2)
       node[midway, above left, xshift=-5pt, yshift=-5pt] {$r_2$};
\node[below left] at (boundary2) {$\partial\Omega$};

\draw[blue, thick, dashed] (a2) circle (1.6);
\node[blue] at (0,2.05) {$B_{r}(\bm a)$};

\draw[->, thick, red] (a2) -- (20:1.6)
      node[midway, above right, xshift=10pt, yshift=-2pt] {$\delta$};

\fill[green!25] plot[smooth cycle] coordinates {
  (-0.8,0.15) (-0.55,0.95) (0.1,1.2) (0.75,0.9) (0.8,-0.15) (0.3,-0.75) (-0.45,-0.7) (-0.85,-0.2)};
\draw[green!60!black, thick] plot[smooth cycle] coordinates {
  (-0.8,0.15) (-0.55,0.95) (0.1,1.2) (0.75,0.9) (0.8,-0.15) (0.3,-0.75) (-0.45,-0.7) (-0.85,-0.2)};
\node[green!50!black] at (0.0,0.6) {$\mathcal L_\gamma$};


\fill (a2) circle (3pt) node[below] {$\bm a$};

\node at (-1,4.5) {\textbf{Case: $r_2 > \delta$ }};
\end{scope}
\end{tikzpicture}%
}

\caption{%
Illustration of Theorem~\ref{thm:lie-decay-region}, Case $c=p+1$. 
Orange $\mathcal X$: $\bm f$ is $\mathcal C^{p+1}$. 
Gray $\Omega\subset\mathcal X$: $\mathcal L_{f_p}V(\bm x)\le0$. 
Blue $B_{r}(\bm a)$: $V$ is a strict local Lyapunov function of $\bm f(\bm x)$. $r_2=\mathrm{dist}(\bm a,\partial\Omega)$. Green $\mathcal L_\gamma$: positively invariant sublevel set component  $\mathcal L_\gamma\subset B_r(\bm a)$ and $\mathcal L_\gamma\subset \mathrm{ROA}(\bm a)$. 
Left $\delta>r_2$, right $r_2>\delta$.}
\label{fig:stability_cases_inputs}
\end{figure}
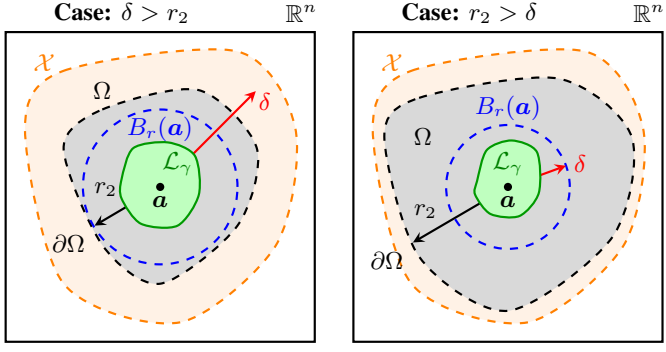

Figure~\ref{fig:stability_cases_inputs} summarizes the main elements of Theorem \ref{thm:lie-decay-region}. Example~\ref{exmp:first_exmp} illustrates this on a system with a finite $\mathrm{ROA}$.

\begin{figure}
    \centering
    \includegraphics[width=0.8\columnwidth]{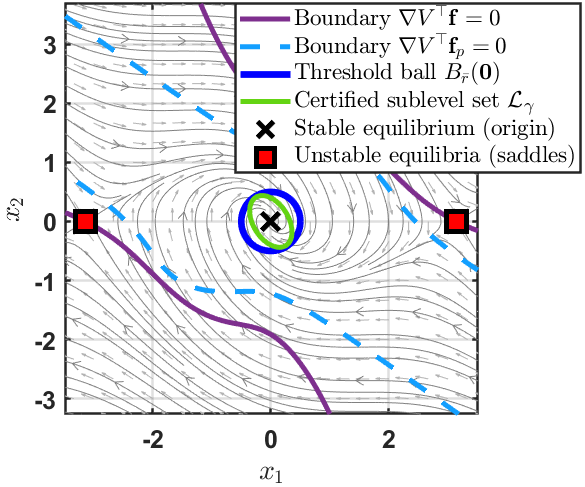}
    \caption{Phase portrait of $\dot{\bm{x}}=\bm f(\bm x)$, certified $\mathcal L_\gamma\subset B_{0.5}(\bm 0)$.}
    \label{fig:phasePort}
\end{figure}

\begin{exmp}[Pendulum with Nonlinear Damping]\label{exmp:first_exmp} Consider a pendulum with nonlinear damping $\tanh(\dot\theta)$, with $x_1=\theta$, $x_2=\dot\theta$, $\bm f(\bm x) = [x_2,\; -\sin(x_1) - \tanh(x_2)]^\top.$
The origin is locally asymptotically stable; saddle equilibria at $(\pm\pi,0)$ bound the $\mathrm{ROA}(\bm 0)$. With $p=3$, using $\sin(x_1)=x_1-\tfrac{1}{6}x_1^3+O(x_1^5)$ and $\tanh(x_2)=x_2-\tfrac{1}{3}x_2^3+O(x_2^5)$:
\[
\bm f_3(\bm x) = \begin{bmatrix} x_2 \\ -x_1 - x_2 + \frac{1}{6}x_1^3 + \frac{1}{3}x_2^3 \end{bmatrix},\quad
\bm \rho_4 = \bm f - \bm f_3.
\]
We take the Lyapunov candidate $V(\bm x)=\tfrac12\bm x^\top P\bm x$, where $P=P^\top\succ0$ solves $A^\top P+PA=-I$ for $A=J\bm f(\bm 0)$, 
\[
A=\begin{bmatrix}0 & 1\\-1 & -1\end{bmatrix} ,\quad
P=\begin{bmatrix}1.500 & 0.500\\ 0.500 & 1.000\end{bmatrix},
\]
where $\lambda_{\min}(P)=0.691,\quad\lambda_{\max}(P)=1.809$. The Lie derivative along $\bm f_3$ is $\mathcal L_{f_3}V(\bm x)=-\tfrac12\|\bm x\|_2^2+\tfrac{1}{12}x_1^4+\tfrac{1}{6}x_1x_2^3+\tfrac{1}{6}x_1^3x_2+\tfrac{1}{3}x_2^4.$ On $\Omega=B_{0.5}(\bm 0)$, the quartic terms satisfy $\left|\tfrac{1}{12}x_1^4 + \tfrac{1}{6}x_1x_2^3 + \tfrac{1}{6}x_1^3x_2 + \tfrac{1}{3}x_2^4\right|
\le \tfrac34 \norm{\bm x}_2^4
\le \tfrac{3}{16}\norm{\bm x}_2^2$, so $\mathcal L_{f_3}V(\bm x)\le-\tfrac{5}{16}\|\bm x\|_2^2\le0$, verifying~\ref{b1}. Condition~\ref{b2} holds since $\nabla V=P\bm x\neq\bm 0$ for $\bm x\neq\bm 0$. For~\ref{dec-2}, choose $c=1$, $\delta=0.5$; since $-\mathcal L_{f_3}V\ge\tfrac{5}{16}\|\bm x\|_2^2$ and $\|\nabla V\|_2\le 1.809\|\bm x\|_2$, condition~\ref{dec-2} holds with $\mu=0.17\le(5/16)/1.809$. To bound the remainder on $\overline{B}_{0.5}(\bm 0)$, we use $\|\nabla^4\bm f(\bm x)\|_{\mathrm{op}}\le\|\nabla^4\bm f(\bm x)\|_{\mathrm{F}}$, where the only nonzero fourth-order entries are $\partial^4 f_2/\partial x_1^4=-\sin(x_1)$ and $\partial^4 f_2/\partial x_2^4=-\tanh^{(4)}(x_2)$, 
\[
\|\nabla^4\bm f(\bm x)\|_{\mathrm{op}}\le\|\nabla^4\bm f(\bm x)\|_{\mathrm{F}}=\sqrt{\!\left(\tfrac{\partial^4 f_2}{\partial x_1^4}\right)^{\!2}\!+\!\left(\tfrac{\partial^4 f_2}{\partial x_2^4}\right)^{\!2}}.
\]
Numerical evaluation of $\max\|\nabla^4\bm f(\bm x)\|_{\mathrm{F}}$ on $\overline{B}_{0.5}(\bm 0)$ yields $m\le 4.1$, hence $k_p\le0.17$ and $r_1\ge1$. Therefore $\bar r=\min\{r_1,\delta,\mathrm{dist}(\bm 0,\partial\Omega)\}=0.5$, and $V$ is a strict Lyapunov function for $\dot{\bm x}=\bm f(\bm x)$ on every $B_r(\bm 0)$, $r\in(0,0.5)$. Since $V$ is quadratic, $\gamma_r^\star=\underline\gamma(r)=\tfrac12\lambda_{\min}(P)r^2=0.3455\,r^2$. With $r=0.49$, $\gamma_r^\star=0.083$, so every $\gamma\in(0,0.083)$ yields an invariant $\mathcal L_\gamma\subset\mathrm{ROA}(\bm 0)$, see Fig.~\ref{fig:phasePort}.
\end{exmp}

\section{Numerical Computation}
\label{sec:POLYNOMIALOPTIMIZATION}

Theorem~\ref{thm:lie-decay-region} provides a rigorous route to certify local asymptotic stability of $\dot{\bm x}=\bm f(\bm x)$ via the polynomial truncation $\dot{\bm x}=\bm f_p(\bm x)$. This allows leveraging polynomial tools \cite{parrilo2000structured} to systematically search for a Lyapunov function for $\bm f_p$ that also certifies the original dynamics $\bm f$. Here we develop a computational pipeline that couples SOS optimization with a grid search to compute a Lyapunov function $V(\bm x)$, maximize the threshold radius $\bar r$, and then extract a certified invariant sublevel set $\mathcal L_\gamma\subset\mathrm{ROA}(\bm a)$ guaranteed by Theorem~\ref{thm:lie-decay-region}.

\begin{defn}[Sum-of-squares (SOS) polynomial]
Let $\bm x\in\mathbb R^n$, and let $p(\bm x)\in\mathbb R[\bm x]$ be a
multivariate polynomial. We say that $p(\bm x)$ is a sum-of-squares (SOS) polynomial, denoted $p(\bm x)\in\Sigma[\bm x]$, if there exist polynomials $q_1(\bm x),\ldots,q_j(\bm x)\in\mathbb R[\bm x]$ such that
$ p(\bm x)=\sum_{i=1}^{j} q_i(\bm x)^2 .$
\end{defn}

Note that if $p(\bm x)\!\in\!\Sigma[\bm x]$, then $p(\bm x)\!\geq\! 0 \; \forall \bm x\!\in\!\mathbb{R}^n$, so we will use a sum-of-squares polynomial as a relaxation for positive polynomials. Checking SOS feasibility is equivalent to solving a Semidefinite Program (SDP) \cite{parrilo2000structured}.

\subsection{SOS Relaxation and Grid Search Optimization}

If $\bm a=\bm 0$, Theorem~\ref{thm:lie-decay-region} requires a polynomial $V(\bm x)$ satisfying
$-\nabla V(\bm x)^\top \bm f_p(\bm x)\ge
\mu \|\nabla V(\bm x)\|_2 \|\bm x\|_2^c$ on $B_\delta(\bm 0)$, which is not
directly convenient for SDP modeling due to the term
$\|\nabla V(\bm x)\|_2$. We then impose the sufficient condition
\begin{equation}\label{eq:sos_decay_cond}
-\nabla V(\bm x)^\top \bm f_p(\bm x)\ge
\epsilon \|\bm x\|_2^{2s},
\qquad \forall\,\bm x\in B_\delta(\bm 0),
\end{equation}
where $\epsilon\!>\!0$ and $s\!\in\!\mathbb N$, $s\!\ge\!1$, are prescribed. If $V$ has a positive definite quadratic part, then there exists $L_\delta\!>\!0$ such that $\|\nabla V(\bm x)\|_2\!\le\! L_\delta\|\bm x\|_2$ $\forall\bm x\!\in \!B_\delta(\bm 0)$, for example with $L_\delta\!=\!\sup_{\bm x\in B_\delta(\bm 0)} \|\nabla^2V(\bm x)\|_{\mathrm{op}}$. Hence \eqref{eq:sos_decay_cond} implies condition~\ref{dec-2} for any $c\in[2s-1,p+1]$ provided $\mu\!\le\! \epsilon/(L_\delta\delta^{c+1-2s})$. In particular, choosing $c\!=\!2s\!-\!1$ gives $\mu\!\le \!\epsilon/L_\delta$. In the numerical examples below, we use $s\!=\!1$, which gives $c=1$. Since the right-hand side of \eqref{eq:sos_decay_cond} is positive for $\bm x\!\neq\!\bm 0$, condition~\ref{b2} also follows automatically on $B_\delta(\bm 0)$.

Since increasing $\delta$ can increase both the Taylor-remainder and
gradient-growth bounds, maximizing $\delta$ does not necessarily maximize
$\bar r$. Thus, we evaluate a grid $\Delta=\{\delta_k\}_{k=1}^{N_\delta}$ and,
for each $\delta_k$, solve the SOS-constrained SDP
\begin{subequations}
\label{eq:sos_feasibility_quartic_generic}
\begin{align}
\min_{V,p,q}\quad
& \operatorname{trace}(P_2(V)) \notag\\
\text{s.t.}\quad
& V(\bm0)=0,\quad \nabla V(\bm0)=\bm0,\quad
P_2(V)\succeq \epsilon_{P2}I,\label{eq:sos_P2_PD_generic}\\
& V-\epsilon_V\|\bm x\|_2^2
-p(\delta_k^2-\|\bm x\|_2^2)\in\Sigma[\bm x],\label{eq:sos_V_PD_generic}\\
& -\nabla V^{\!\top\!} \bm f_p
-\epsilon\|\bm x\|_2^{2s}
-q(\delta_k^2-\|\bm x\|_2^2)\in\Sigma[\bm x],\label{eq:sos_Vdot_ND_generic}\\
& p,q\in\Sigma[\bm x],
\end{align}
\end{subequations}
where $\epsilon_V,\epsilon,\epsilon_{P2}>0$ are fixed for simplicity and normalization, the polynomial degrees are chosen appropriately,
and $P_2(V)$ is defined by $V_2(\bm x)=\bm x^\top P_2(V)\bm x$. For each feasible $\delta_k$, the threshold radius and the sublevel set are found using Theorem~\ref{thm:lie-decay-region}. This procedure is shown in Algorithm~\ref{alg:sos_grid_sublevel}, with $c<p+1$. The exponent $c$ specifies the order at which the truncated-model decay dominates the Taylor remainder. Although larger $c$ can capture higher-order decay, it does not necessarily enlarge the certified radius, since $r_1=(\mu/k_p)^{1/(p+1-c)}$ and, under the SOS decay condition, $\mu=\epsilon/(L_\delta\delta^{c+1-2s})$. Thus larger $c$ may reduce the admissible $\mu$, while the case $c=p+1$ additionally requires $\mu>k_p$.

\begin{algorithm}[t]
\caption{SOS grid search and $\mathrm{ROA}$-sublevel set}
\label{alg:sos_grid_sublevel}
\begin{algorithmic}[1]
\STATE \textbf{Input:} $\Delta=\{\delta_k\}_{k=1}^{N_\delta}$, order $p$, decay parameters $s,c$.
\STATE \textbf{Initialize:} $\bar r_{\mathrm{best}}=0$.
\FOR{each $\delta_k\in\Delta$}
    \STATE Solve \eqref{eq:sos_feasibility_quartic_generic} on $B_{\delta_k}(\bm0)$ to obtain $V_k$.
    \STATE \textbf{if infeasible, continue}
    \STATE Compute $L_{\delta_k}$ s.t. $\|\nabla V_k(\bm x)\|_2\!\le\! L_{\delta_k}\|\bm x\|_2$ on $B_{\delta_k}(\bm0)$.
    \STATE Compute
    $m_{\delta_k}=\max_{\overline B_{\delta_k}(\bm0)}
    \|\nabla^{p+1}\bm f\|_{\mathrm{op}}$ and
    $k_{p,k}=m_{\delta_k}/(p+1)!$.
    \STATE Set $\mu_k=\epsilon/(L_{\delta_k}\delta_k^{c+1-2s})$.
    \STATE Set $r_{1,k}=(\mu_k/k_{p,k})^{1/(p+1-c)}$, $\bar r_k=\min\{r_{1,k},\delta_k\}$.
    \IF{$\bar r_k>\bar r_{\mathrm{best}}$}
        \STATE Store $\bar r_{\mathrm{best}}=\bar r_k$, $\delta^*=\delta_k$, and $V^*=V_k$.
    \ENDIF
\ENDFOR
\STATE Set $r=0.99\,\bar r_{\mathrm{best}}$.
\STATE Compute $\hat\gamma_r$, either as
$\gamma_r^\star=\min_{\|\bm x\|_2=r}V^*(\bm x)$ numerically or as
$\underline\gamma(r)$ from Theorem~\ref{thm:lie-decay-region}.
\STATE Choose any $\gamma$ such that $0<\gamma<\hat\gamma_r$.
\STATE \textbf{return} $V^*$, $\delta^*$, $\bar r_{\mathrm{best}}$, $r$, $\gamma$,
$\mathcal L_\gamma=\{\bm x\in\Omega:V^*(\bm x)\le\gamma\}$.
\end{algorithmic}
\end{algorithm}

\begin{figure}
    \centering
    \includegraphics[width=0.74\columnwidth]{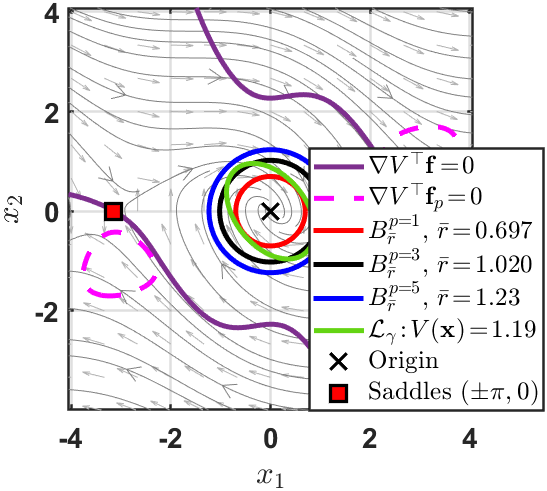} 
    \caption{Certified threshold balls for $p=1,3,5$, Example~\ref{exmp:sos_grid_quartic}.}
    \label{fig:phasePort_sos}
\end{figure}

\begin{exmp}[Maximal Threshold Radius by Grid Search]\label{exmp:sos_grid_quartic}
We apply Algorithm~\ref{alg:sos_grid_sublevel} to the pendulum of Example~\ref{exmp:first_exmp}, using its $p=5$ polynomial truncation $\bm f_p(\bm x)$ and searching for a quadratic Lyapunov function $V(\bm x)$. We set $\epsilon=1$, $\epsilon_V=\epsilon_{P2}=10^{-6}$, $s=1$, and $c=1$. The SOS-constrained SDP \eqref{eq:sos_feasibility_quartic_generic}, with $\deg(V)=2$, $\deg(p)=2$, and $\deg(q)=4$, is solved on a grid $\delta_k\in[0.5,3]$ of $30$ points using MOSEK~\cite{mosek} via YALMIP~\cite{Lofberg2009}, where each point of the grid took $0.15$ s to run in a 12th Gen Intel(R) Core(TM) i7-12700H at 2.30GHz. Following Algorithm~\ref{alg:sos_grid_sublevel}, the best certified value is obtained at $\delta^*=1.276$, $L_{\delta^*}=4.775,$ $m_{\delta^*}=52.789.$ Thus $\mu=1/L_{\delta^*}=0.209$, $k_p=m_{\delta^*}/6!=0.0733$, and $ r_1=1.23,$ $\bar r_{\mathrm{best}}=\min\{r_1,\delta^*\}=1.23.$

Therefore, $V^*(\bm x)$ is a strict Lyapunov function of $\dot{\bm x}=\bm f(\bm x)$ on every ball $B_r(\bm0)$ with
$r\in(0,\bar r_{\mathrm{best}})$. For visualization, we choose $r=0.99\bar r_{\mathrm{best}}=1.2212$, which gives
$\gamma_r^\star=1.4954$; hence we take $\gamma=1.1963<\gamma_r^\star$.
Fig.~\ref{fig:phasePort_sos} shows the phase portrait and compares the certified balls for $p=1$, $p=3$, and $p=5$, with $\bar r=0.697$, $1.020$, and $1.23$, respectively. For the reported $p=5$ certificate, the displayed connected sublevel set
$\mathcal L_\gamma=\{\bm x\in\Omega:V^*(\bm x)\le\gamma\}\subset B_r(\bm0)$
is positively invariant and satisfies
$\mathcal L_\gamma\subset\mathrm{ROA}(\bm0)$.
\end{exmp}

\subsection{Analysis for Closed Loop Feedback Control Systems}

The stability transfer result extends directly to control systems $\dot{\bm x}\!=\!\bm f(\bm x,\bm u)$ under a fixed state-feedback law $\bm u\!=\!\bm \kappa(\bm x)$, since the closed loop $\bm f_{cl}(\bm x)\!=\!\bm f(\bm x,\bm \kappa(\bm x))$ is autonomous. In this case, the SOS pipeline of Section~\ref{sec:POLYNOMIALOPTIMIZATION} can then be applied to compute a Lyapunov function $V(\bm x)$, a radius $\bar r$, and a certified invariant set $\mathcal L_\gamma\!\subset \!\mathrm{ROA}(\bm a)$. In contrast, jointly optimizing $\bm \kappa(\bm x)$ and $V(\bm x)$ (both polynomials) introduces bilinear constraints and leads to a substantially harder nonconvex problem \cite{tan_packard_clf_sos,jarvis2003some,tan2008stability}.

We therefore use the framework as a certification-oriented tuning tool over a prescribed LQR family. Given a target initial condition $\bm x_0$, we search over a finite grid of LQR weights $R$ and select the largest certified value in that grid, corresponding to the lowest-gain controller among the tested LQR designs, such that the nonlinear closed loop admits a certified invariant set $\mathcal L_\gamma\subset \mathrm{ROA}(\bm a)$ containing $\bm x_0$. This procedure is summarized in Algorithm~\ref{alg:min_effort_lqr_sublevel}, where we assume that the relevant sublevel sets are connected ($S_\gamma=\mathcal L_\gamma$), otherwise one needs to check that $\bm x_0\in\mathcal L_\gamma$.

\begin{algorithm}[t]
\caption{Minimum-effort LQR certification}
\label{alg:min_effort_lqr_sublevel}
\begin{algorithmic}[1]
\STATE \textbf{Input:} $\bm x_0$, grid $\{R_k\}$, $Q\!\succeq\!0$, order $p$, SOS parameters.
\FOR{each $R_k$}
    \STATE Get LQR gain $K_{R_k}$ and form $\bm f(\bm x,-K_{R_k}\bm x)$.
    \STATE Construct the $p$th-order Taylor model $\bm f_{cl,p}(\bm x)$.
    \FOR{each candidate $\delta$ in a prescribed sequence}
        \STATE Solve \eqref{eq:sos_feasibility_quartic_generic} on $B_\delta(\bm 0)$ for $V(\bm x)$.
        \STATE \textbf{if infeasible, continue}
        \STATE Compute $\mu$, $k_p$, and $\bar r$ from Theorem~\ref{thm:lie-decay-region}.
        \FOR{each $\eta$ in a prescribed set, e.g.\ $\eta=0.99$}
            \STATE Set $r=\eta\,\bar r$ and estimate $\gamma_r^\star  =\min_{\|\bm x\|_2=r}V(\bm x)$.
        \ENDFOR
        \STATE Let $\hat \gamma_r$ be the largest $\gamma_r^\star$ over the tested $\eta$.
        \IF{$V(\bm x_0)<\hat \gamma_r$}
            \STATE Choose any $\gamma$ such that $V(\bm x_0)<\gamma<\hat \gamma_r$.
            \STATE \textbf{return} $R_k$, $K_{R_k}$, $V$, $r$, $\gamma$.
        \ENDIF
    \ENDFOR
\ENDFOR
\STATE \textbf{return} no controller in the prescribed grid certifies $\bm x_0$.
\end{algorithmic}
\end{algorithm}

If the procedure succeeds, it returns an LQR gain that certifies $\bm x_0\in\mathrm{ROA}$ for the true closed loop dynamics, while selecting (within a LQR family) a minimum-effort controller computed from the linearized model. It is important to note that the stability guarantee for $\bm x_0$ is with respect to the nonlinear closed loop system, while the optimal effort is with respect to the linearized system (LQR infinite horizon).

\begin{figure}
\centering
\includegraphics[width=0.8\linewidth]{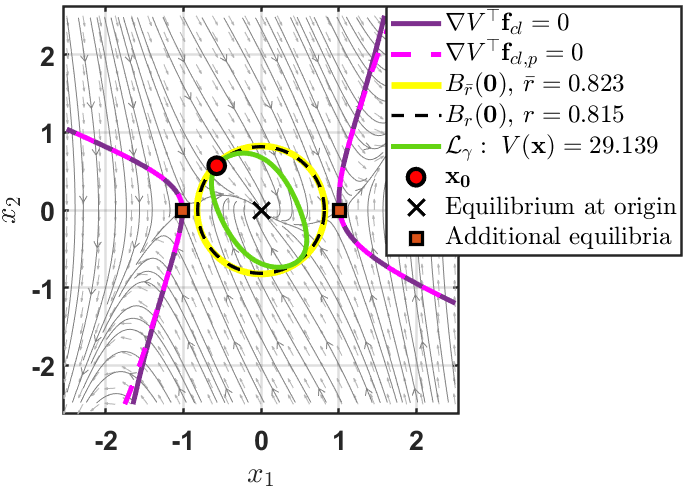}
\caption{Example~\ref{exmp:min_effort_lqr} phase portrait with $\bm x_0\in \mathcal L_\gamma\subset \mathrm{ROA}(\bm 0)$.}
\label{fig:phase_portraits_comparison}
\end{figure}

\begin{exmp}[Reduced Effort LQR with Initial Condition]\label{exmp:min_effort_lqr}
Consider the nonpolynomial control system $\dot{\bm x} = [ x_2,\; x_1 + x_1^3 - 0.1 \tanh(x_2) + u ]^\top.$
The origin is an unstable equilibrium for $u=0$. We aim to find the LQR controller $\bm u = -K_R \bm x $, obtained by solving the algebraic Riccati equation with $Q=I$, that uses the largest scalar $R$ while ensuring $\bm x_0 = [-0.57, 0.57]^\top\!\in\mathrm{ROA}(\bm 0)$.
We use the $p=5$ Taylor approximation $\bm f_{cl,p}(\bm x)$ of the closed loop system $\dot{\bm x} = \bm f(\bm x, -K_R \bm x)$. We search for a quartic Lyapunov function $V(\bm x)$ using the relaxation \eqref{eq:sos_decay_cond} with $\epsilon=1.0$, $s=1$.

Following Algorithm~\ref{alg:min_effort_lqr_sublevel}, searching over $R \in [1, 100]$ (logarithmically spaced, 25 points, from largest to smallest), we find $R^* = 21.544$ that certifies the initial condition $\bm x_0$, where each point took around $0.18$ s to run in a 12th Gen Intel(R) Core(TM) i7-12700H at 2.30 GHz. The corresponding gain is $K_{R^*} = [2.0229, 1.9254]$. For this controller, the SOS problem finds a valid quartic $V(\bm x)$ and $\delta = 0.9673$ such that $V(\bm x_0)=29.132<\gamma<\hat\gamma_r$, with $\gamma=29.139$, then the initial condition $\bm x_0$ is certified to be within the $\mathrm{ROA}(\bm 0)$.

Figure~\ref{fig:phase_portraits_comparison} shows the system stabilized by the grid-optimized controller $K_{R^*}$. The plot confirms the target initial condition $\bm x_0$ belongs to $\mathcal L_\gamma\subset \mathrm{ROA}(\bm 0)$. This result yields two main benefits: a 16.53\% reduction in control gain norm (vs. $R=1$), and critically, a formal certificate that $\bm x_0$ is within the $\mathrm{ROA}(\bm 0)$ for the nonpolynomial closed loop system, a guarantee that standard LQR design cannot provide alone.
\end{exmp}

\section{Conclusions}\label{sec:DiscussionConclusions}

We developed a stability transfer result that certifies local asymptotic stability of a nonpolynomial system from a Lyapunov certificate for a finite-order Taylor truncation. Unlike classical Jacobian analysis, the proposed result does not rely on hyperbolicity and yields an explicit threshold radius with a computable invariant sublevel set approximation of the $\mathrm{ROA}$.

We also presented an SOS-based computational pipeline to search for polynomial Lyapunov functions, maximize the threshold radius, and extract certified invariant sublevel sets. The examples show how the method certifies initial conditions inside the $\mathrm{ROA}$ of the original nonpolynomial closed-loop
system. A related application is SOS-based nonlinear controller synthesis for spacecraft relative motion \cite{ACC_mine}. Integrating the present stability-transfer result into that setting would provide explicit local $\mathrm{ROA}$ guarantees for controllers designed from polynomial or lifted
approximations of nonpolynomial orbital dynamics.
 
\bibliographystyle{IEEEtran}
\bibliography{Ref}

\end{document}